\documentclass[preprint]{aastex631}

\shorttitle{Implications from Jupiter Gas-Driven Migration}
\shortauthors{Deienno et al.}
\graphicspath{{./}{figures/}}

\begin{document}

\title{Implications of Jupiter Inward Gas-Driven Migration for the Inner Solar System.}

\author[0000-0001-6730-7857]{Rogerio Deienno}
\affiliation{Department of Space Studies, Southwest Research Institute, 1050 Walnut Street, Suite 300, Boulder, CO 80302, USA}

\author[0000-0003-1878-0634]{Andr\'e Izidoro}
\affiliation{Department of Physics and Astronomy 6100 MS 550, Rice University, Houston, TX 77005, USA}

\affiliation{Department of Earth, Environmental and Planetary Sciences, 6100 MS 126, Rice University, Houston, TX 77005, USA}


\author{Alessandro Morbidelli}
\affiliation{D\'epartement Lagrange, University
of Nice–Sophia Antipolis, CNRS, Observatoire de la C\^ote d'Azur, Nice, France.}

\author[0000-0002-4547-4301]{David Nesvorn\'y}
\affiliation{Department of Space Studies, Southwest Research Institute, 1050 Walnut Street, Suite 300, Boulder, CO 80302, USA}

\author[0000-0002-1804-7814]{William F. Bottke}
\affiliation{Department of Space Studies, Southwest Research Institute, 1050 Walnut Street, Suite 300, Boulder, CO 80302, USA}



\begin{abstract}

The migration history of Jupiter in the sun's natal disk remains poorly constrained. Here we consider how Jupiter's migration affects small-body reservoirs and how this constrains its original orbital distance from the Sun. We study the implications of large-scale and inward radial migration of Jupiter for the inner solar system while considering the effects of collisional evolution of planetesimals. We use analytical prescriptions to simulate the growth and migration of Jupiter in the gas disk. We assume the existence of a planetesimal disk inside Jupiter's initial orbit. This planetesimal disk received an initial total mass and size-frequency distribution (SFD). Planetesimals feel the effects of aerodynamic gas drag and collide with one another, mostly while shepherded by the migrating Jupiter. Our main goal is to measure the amount of mass in planetesimals implanted into the main asteroid belt (MAB) and the SFD of the implanted population. We also monitor the amount of dust produced during planetesimal collisions. We find that the SFD of the planetesimal population implanted into the MAB tends to resemble that of the original planetesimal population interior to Jupiter. We also find that unless very little or no mass existed between 5 au and Jupiter's original orbit, it would be difficult to reconcile the current low mass of the MAB with the possibility that Jupiter migrated from distances beyond 15 au. This is because the fraction of the original disk mass that gets implanted into the MAB is very large. Finally, we discuss the implications of our results in terms of dust production to the so-called {\it NC-CC isotopic dichotomy}. 

\end{abstract}



\section{Introduction} \label{sec:intro}

In order to accrete substantial amounts of atmosphere and become gas giants, planets like Jupiter must form while gas in the protoplanetary disk is still around. During the process of growth and gas accretion, disk-planet tidal interactions may result in planet gas-driven migration \citep[e.g.,][]{Baruteau2013,Nelson2018}. Although disk-planet tidal interactions are very complex, a simplified summary is that planet gas-driven migration can be classified into two main modes, Type-I and Type-II. These two modes of migration mostly depend on the characteristics of the protoplanetary disk and planet mass. Planets that are not large enough to trigger runaway gas accretion may experience large-scale Type-I migration, which is mostly inward \citep[e.g.,][and references therein]{Bitsch2019}. On the other hand, planets that are massive enough to trigger runaway gas accretion may disturb the protoplanetary disk gas surface density and carve a deep gap. In this situation, the planet experiences reduced migration in the Type-II mode \citep[e.g.,][which depends on the disk viscosity]{Ndugu2021,Schneider2021a,Schneider2021b}.

Planet migration seems to be a generic process of planet formation. The existence of systems of hot super-Earths\footnote{Planets with sizes between 1 and 4 Earth radii, orbital periods of less than 100 days, and masses larger than that of the Earth but smaller than Neptune.} with multiple planet pairs near or in mean-motion resonances (MMRs) strongly suggests that Type-I migration occurred during the formation of these planetary systems \citep{Izidoro2017,Izidoro2021}.  Similarly, the existence of cold Jupiter\footnote{Planets with masses larger than 0.3 Jupiter masses and orbiting the star at distances larger than 1 au.} planets in systems composed of inner super-Earths \citep{Zhu2018} suggest that cold Jupiters likely stopped migration before reaching distances smaller than 1 au. Yet cold Jupiters tend to migrate over large orbital distances during their formation while in Type-I mode \cite[e.g.,][]{Bitsch2015b}. The length of inward migration is not constrained. Forming a giant planet farther out from the star may sound appealing. This is because Type-II migration, although slower than Type-I migration, can still bring a planet from regions more distant than 10 au to the closest regions of the disk during the disk lifetime. Once forming farther out, e.g. $a>$ 20 au, it takes longer to migrate all the way to the star. Therefore, this view has been invoked as a potential way to explain the occurrence of cold Jupiters \citep[e.g.,][]{Bitsch2019}. 

This suggests that Jupiter in our solar system potentially formed farther away from the Sun and migrated inward \citep[e.g.,][]{Johansen2017}. Modeling of Jupiter's inward migration often follows a proposition known as {\it planetary growth tracks}, i.e., analytic predicted paths in the mass-orbital radius diagram that a planet would follow once accounting for planetesimal and pebble accretion along with atmospheric growth and disk-planet tidal interactions. Still, there are no existing constraints in regard to how much farther from the Sun Jupiter could have formed. Placing such constraints is difficult given the uncertainties about the original protoplanetary disk. Indeed, Jupiter's current orbit can be reproduced from many different initial configurations as a result of small changes in the disk's parameters, e.g., gas column density, viscosity, and fraction of initial solids (planetesimals and pebbles), among other quantities \citep[e.g.,][]{Bitsch2015b, Johansen2017,Johansen2019}. On the other hand, it is well known that Jupiter's inward migration would largely affect the small-body population interior to its orbit \citep[e.g.][]{Walsh2011,Batygin2015,Raymond2017a,Carter2020}.

In this work, we aim to find constraints on Jupiter's original orbital distance from the Sun. For that, we consider Jupiter's core growing at different distances from the Sun and follow its inward gas-driven migration via {\it planetary growth tracks}. A massive disk of planetesimals is placed interior to the orbit of Jupiter. During the gas disk phase, planetesimals are caught in MMR and transported by the migrating Jupiter. We follow the collisional evolution of such a planetesimal population. Colliding planetesimals often fragment and evolve into small (subkilometer-sized) debris or even become dust \citep{Batygin2015,Deienno2020}, especially once in MMRs. 
Sufficiently small planetesimals and dust grains experience fast orbital decay \citep{Weidenschilling1977} due to strong aerodynamic gas drag. They can reach the inner regions of the solar system and eventually be lost. Our constraints are then based primarily on the relationship between Jupiter's original orbit and the total mass transported to the inner solar system (i.e., the main asteroid belt, MAB, and terrestrial planet region). Specifically, we compare the amount of mass implanted into the MAB with the current MAB mass \citep[$\approx$ 5$\times$10$^{-4}~M_{\oplus}$;][]{DeMeo2013,DeMeo2014}. We also measure the amount of mass in dust and pebble-sized fragments\footnote{In this work, we define pebbles as particles with sizes in the range 1 mm $\le r_{planetesimal} <$ 1 m and dust objects with $r_{planetesimal} <$ 1 mm; see Section \ref{sec:model}} that will reach the terrestrial region and evaluate their contribution to Earth and Mars. Finally, as the planetesimal population is allowed to collisionally evolve in our simulations, we follow the size-frequency distribution (SFD) of the MAB implanted population. We use this implanted SFD to discuss implications for the SFD of the original planetesimal population interior to Jupiter.

We organize this work as follows. In section \ref{sec:model}, we present our model. In section \ref{sec:mab}, we address our primary goal by discussing the implications of Jupiter's inward gas-driven migration to the MAB. In section \ref{sec:sfd}, we discuss implications related to the implanted SFD. In section \ref{sec:nccc}, we discuss implications from the amount of mass in dust and pebble-sized fragments reaching the terrestrial planet region. Section \ref{sec:timing} is devoted to discussing implications related to the timing of Jupiter's formation and its migration speed. Section \ref{sec:concl} summarizes our conclusions.

\section{Model} \label{sec:model}

We added analytical prescriptions for {\it planetary growth tracks} \citep{Johansen2017} to the Lagrangian Integrator for Planetary Accretion and Dynamics (LIPAD) code \citep{Levison2012} to simulate Jupiter's growth and migration. LIPAD is a particle-based (i.e., Lagrangian) code that can follow the collisional/accretional/dynamical evolution of a large number of subkilometer objects through the entire growth process to become planets. It uses the concept of tracer particles to represent a large number of small bodies with roughly the same orbit and size. As described by \cite{Levison2012}, LIPAD is unique in its ability to accurately handle the mixing and redistribution of material due to gravitational encounters, including resonant trapping, while also following the fragmentation and accretional growth of bodies. LIPAD has a prescription of the gaseous nebula from \cite{Hayashi1985}. This gas disk provides aerodynamic drag, eccentricity, and inclination damping on planetesimals and planets. The collisional routines from LIPAD follow \cite{Benz1999} disruption laws. LIPAD is a well-tested code that has been successfully employed in previous studies of the collisional evolution of centimeter- to kilometer-sized planetesimals interacting with planets \citep{Kretke2014,Levison2015a,Levison2015b,Walsh2016,Walsh2019,Deienno2019,Deienno2020,Voelkel2021a, Voelkel2021b, Izidoro2022}, making it ideal for our study. 

In our simulations, Jupiter follows the {\it planetary growth track} prescriptions while interacting with planetesimals \citep[represented by tracer particles; see][for details about tracer particles]{Levison2012}. Planetesimals are allowed to collisionally evolve (grow or fragment) once interacting with Jupiter, among themselves, and once transported via MMRs. We tracked the entire collisional cascade all the way to the point where the planetesimals break into pebble sizes (here defined as particles with sizes in the range 1 mm $\le r_{planetesimal} <$ 1 m)
or become dust particles
(here defined as $r_{planetesimal} <$ 1 mm;  these are removed from the simulation). 
We consider our planetesimals to be made of ice \citep{Benz1999}. This is because we are assuming that they formed beyond the water-ice line, which is predicted to be interior to 5 au \citep{Lambrechts2014,Drazkowska2017} by the time that outer solar system planetesimals formed \citep[$\approx$ 0.5--1 Myr after calcium-aluminium-inclusion -- CAI;][]{Lichtenberg2021,Izidoro2022,Morbidelli2022}.

Jupiter's {\it planetary growth tracks} were designed such that Jupiter would take 1 Myr to grow from a Moon size to its current mass, once migrating from its original location to 5.2 au. The inner edge of our initial planetesimal disk is set to be at 5 au. We have placed the inner edge at 5 au to avoid an overlap with the MAB. The outer edge is at 2 au within Jupiter's initial position. The reason for this is that we are forcing Jupiter to follow analytical expressions of {\it planetary growth track}. This is necessary so that the influence from (or encounters with) nearby massive portions of the planetesimal disk (especially once Jupiter is only a Moon-ish-sized object) does not deviate/scatter the planet away from its predicted {\it planetary growth track}. Our planetesimal disk assumptions allow us to focus solely on the effects that a growing and inward-migrating Jupiter has on planetesimals interior to its orbit. 

\begin{figure}[h!]
    \centering
    \includegraphics[width=\linewidth]{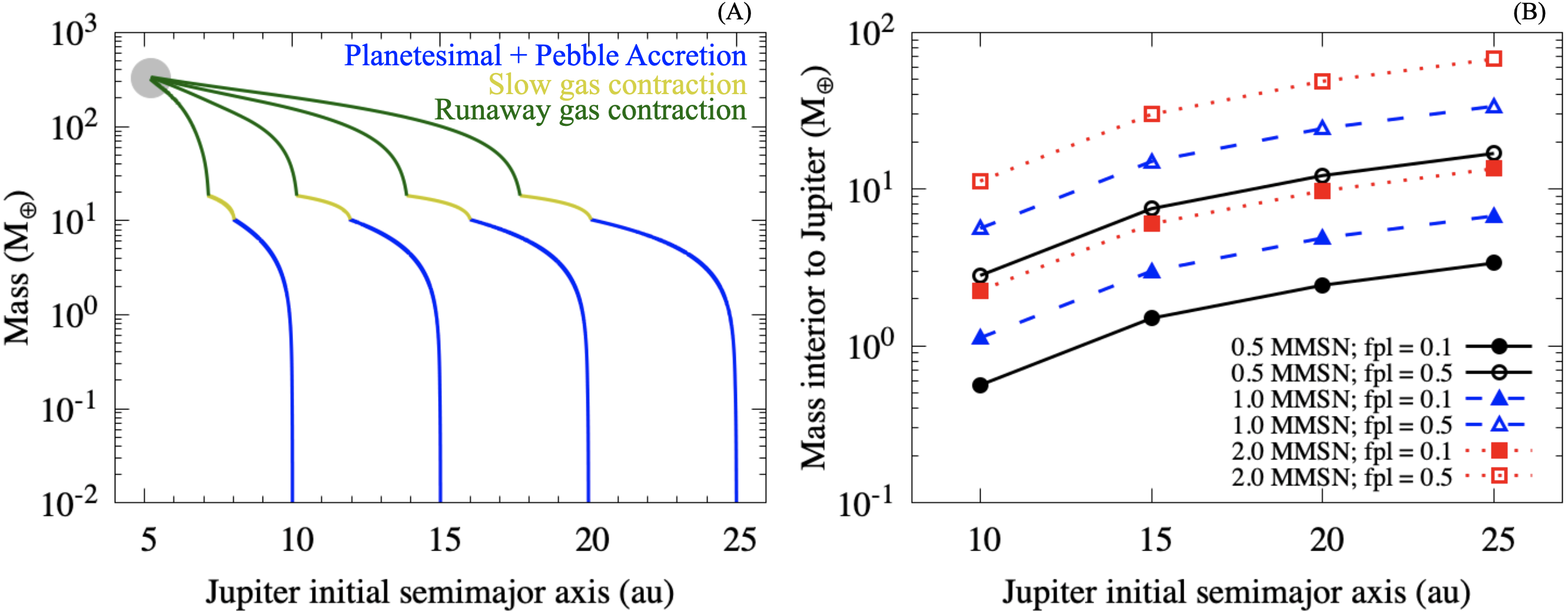}
    \caption{Left (A): {\it Planetary growth tracks} for Jupiter starting at 10, 15, 20, and 25 au as a Moon-sized object (M$_{jup}^{init}$ = 0.01 M$_{\oplus}$) and migrating all the way to 5.2 au (current position with its current mass, M$_{jup} \approx$  330 M$_{\oplus}$; gray circle). Right (B): total mass in the disk interior to Jupiter's initial orbit and outside 5 au (see main text). Symbols in the bottom right represent the many cases studied (radial mass distributions, MMSN, and the initial fraction of the total solid MMSN mass transformed into planetesimals in each given disk, fpl, when assuming a 1\% stellar metallicity).}
    \label{Fig1}
\end{figure}

In order to understand the role of Jupiter’s growth and migration on the mass delivery to the inner solar system, we followed four different {\it planetary growth tracks} for Jupiter: $a_{Jup}^{init}=$ 10, 15, 20, and 25 au (Figure \ref{Fig1}A). We used the same parameters provided by \cite[see their section 5.3]{Johansen2017} while varying the gas column density parameter ($f_g$) in the range 0.2 $\le f_g <$ 0.3 \citep[see also][]{Bitsch2015a,Johansen2019}. We started Jupiter's growth from a Moon-sized object (M$_{jup}^{init}$ = 0.01 M$_{\oplus}$, where M$_{\oplus}$ denotes one Earth mass).

For each growth track, we assumed that a given amount of planetesimals would form interior to Jupiter's orbit and exterior to 5 au (near Jupiter's current orbit).
For simplicity, the initial distribution of planetesimals in the disk follows the minimum-mass solar nebula \citep[MMSN;][]{Hayashi1981} radial density profile. We assumed a smooth distribution ($\Sigma = \Sigma_0 r^{-1}$) and a stellar metallicity of 1\% to determine the total mass in solids. Only a fraction of that total solid mass was converted into planetesimals (fpl = 0.1 and 0.5). Here $\Sigma_0$ in $\Sigma = \Sigma_0 r^{-1}$ is defined as $\zeta \Sigma_{1au}$, where $\Sigma_{1au}$ represents the MMSN surface density at 1 au, equal to 1700 ${\rm g cm^{-2}}$ \citep{Hayashi1981}. We considered three different values for $\Sigma_0$ with $\zeta$ = 0.5, 1, and 2\footnote{Because we already have overlap of total disk masses (see Figure \ref{Fig1}B) once varying fpl and $\Sigma_0$ in the cases studied, performing simulations with intermediate cases (more values of fpl and $\Sigma_0$) would not necessarily improve or change the results and conclusions.} (Figure \ref{Fig1}B). 

The planetesimal population was given an initial cumulative SFD that follows $N(>D) \propto D^{-q}$, where $q=$ 5, with the largest object having $D =$ 1000 km ($\approx$ Ceres) and the smallest $D =$ 100 km. This SFD represents the typical size range of planetesimals formed by the streaming instability \citep[e.g.,][]{Youdin2005}. A slope of $q=$ 5 is also suggested as a good representative of the Kuiper belt population based on observations from the larger objects \citep{Fraser2014}. Such an SFD is also consistent with the proposition that newborn planetesimals have the characteristic size $D \approx$ 100 km, thus being the most numerous \citep{Morbidelli2009,Klahr2020,Klahr2021}. We choose the Kuiper belt instead of the MAB as a first approximation for the initial SFD because the MAB is known to be heavily collisionally evolved \citep{Bottke2005}. This is not the case for the cold population of the Kuiper belt given the large number of observed 100 km class equal-sized binaries \citep{Fraser2017}. Still, we are not claiming that our choice is correct or ultimate. We will discuss the implications of our choice of $q$ in Section \ref{sec:sfd}.

\section{Implications for the \it MAB} \label{sec:mab}

The current MAB presents a total mass of order 10$^{-4}$ of an Earth mass \citep[$\approx$ 5$\times$10$^{-4}~M_{\oplus}$;][]{DeMeo2013,DeMeo2014}. 
This mass is dominated by two main taxonomic types of asteroids, S- and C-types \citep{Gradie1982}. The S-types are water-poor asteroids and thought to have originated at radial distances interior to the water-ice line. The C-type asteroids are water-rich and formed beyond the water-ice line (beyond 5 au in our case; Section \ref{sec:model}). The latter currently represent about 3/4 of the total MAB mass \citep{Mothe-Diniz2003}, i.e., M$_{\rm {C-type}}$ $\approx$ 3.75$\times$10$^{-4}~M_{\oplus}$. We use the total mass of the C-types in the asteroid belt as a constraint. 

In our modeling, we compare M$_{\rm {C-type}}$ with the amount of mass implanted into the MAB during Jupiter's inward gas-driven migration from $a_{Jup}^{init}$ (Section \ref{sec:model}) to 5.2 au. If too much mass in planetesimals is implanted during the gas phase, a subsequent dynamical process happening after gas disk dispersal is necessary to deplete the excess mass. After gas disk dispersal, the only large event capable of such is the giant planet instability \citep[e.g.,][]{Nesvorny2012,Deienno2017,Clement2018,Ribeiro2020}. 
Dynamical models of the giant planet instability \citep[e.g.,][]{Nesvorny2018a,Deienno2018,Clement2019,Nesvorny2021} are capable of producing lower and upper limits of mass depletion in the MAB between 50\% and 99.9\%\footnote{Giant planet evolution leading to more than 99.9\% depletion results in a final orbital excitation and separation of Jupiter and Saturn that are orders of magnitude above the observed values \citep[see cases 4, 7, and 7a in both Figure 2 and Table 2 from][]{Clement2019}.}, respectively. After the giant planet instability, the remaining MAB mass is depleted by an additional factor of 50\% over 4 Gyr due to chaotic diffusion \citep{Minton2010,Deienno2016,Deienno2018}. Therefore, if more than $\approx$ 0.75 M$_{\oplus}$\footnote{M$_{\rm {C-types}} =$ 3.75$\times$10$^{-4}~M_{\oplus}$ = [(M$_{\rm {Implanted}}$ - 99.9\%) - 50\%] M$_{\oplus}$, where M$_{\rm {Implanted}}$ = 0.75 M$_{\oplus}$.} in planetesimals from beyond the water-ice line (C-type) get implanted in the MAB, we should have a more massive MAB today. We report on combinations of $a_{Jup}^{init}$ and the amount of planetesimals interior to Jupiter's orbit that satisfy the aforementioned implantation mass constraint. With that, we aim to constrain the farthest location from the Sun where Jupiter could likely have formed.

\begin{figure}[t]
    \centering
    \includegraphics[width=\linewidth]{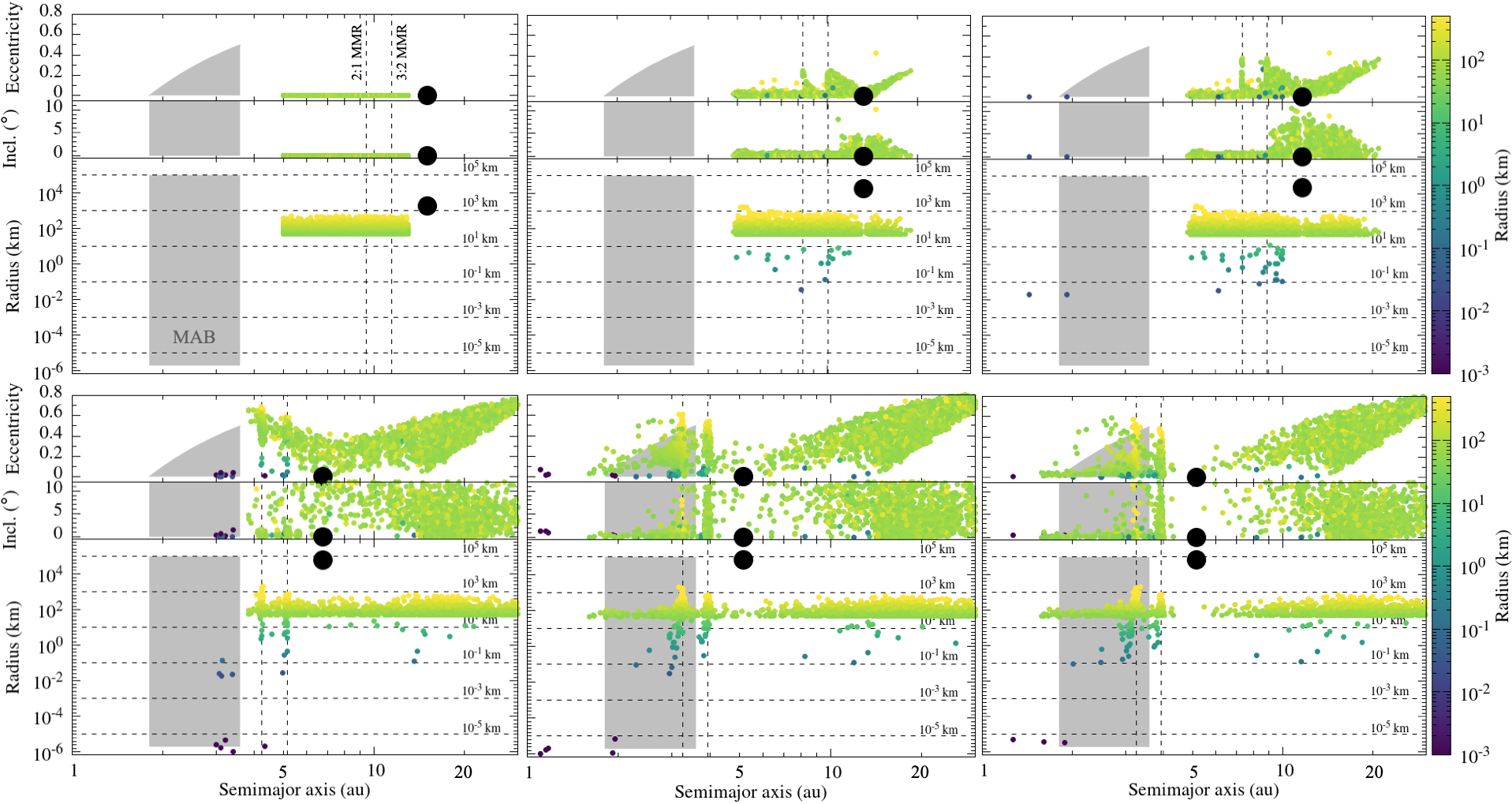}
    \caption{Representation of the implantation mechanism via MMR transport during Jupiter's inward gas-driven migration (case 2.0 MMSN, fpl = 0.5, $a_{Jup}^{init}=$ 15 au). The evolution of the eccentricity, inclination, and planetesimal radii (top, middle, and bottom panels in each plot) as a function of semimajor axis is provided. The planetesimal sizes are colored to improve visualization. Time evolves from left to right and top to bottom, as Jupiter (black filled circle), in this example, moves from $a_{Jup}^{init}=$ 15 au to $a_{Jup}^{final}=$ 5.2 au. Planetesimals are more efficiently transported via 2:1 and 3:2 MMR shepherding (vertical dashed lines). The figure also shows that collisional evolution happens mainly once planetesimals are in MMRs (see Section \ref{sec:sfd} for discussion). The MAB space (gray shaded area) is defined as in \citet[region limited within 1.8 au $\leq a \leq$ 3.6 au]{Deienno2016,Deienno2018}. The horizontal lines in the bottom panels are shown for reference sizes of r = 10$^{-5}$ km, 10$^{-3}$ km, 10$^{-1}$ km, 10$^{1}$ km, 10$^{3}$ km, and 10$^{5}$ km (bottom to up).}
    \label{Fignew}
\end{figure}

In Figure \ref{Fignew}, we present how the implantation mechanism occurred during Jupiter's inward gas-driven migration. The main mechanism of implantation is MMR shepherding, and the most important MMRs for such are the 2:1 and 3:2. In this work, we assume that Jupiter's migration would end either once Jupiter carves a very deep gap in the nebula (not modeled) or once the gas in the disk is almost fully dispersed (we assumed that the gas disk would exponentially disperse within a 2 Myr timescale). With at least a small component of the gas still surviving in the disk, we have that larger planetesimals remain near the resonances that shepherded them with eccentricities and inclinations not effectively damped by the gas. Smaller planetesimals, on the other hand, due to their stronger interaction with the remnant gas component of the dispersing disk, would be quickly pushed away from the resonant equilibrium and released from the MMRs that shepherded them. The amount of orbital damping for small planetesimals depends mostly on how much gas would still be left in the disk. This can have large implications, especially on the semimajor axis distribution of the implanted C-type planetesimals into the MAB. However, because we do not know how much gas can be left in the disk after Jupiter stops migrating, we prefer not to speculate further about these issues in the present work.

There is no dependence between implantation and initial eccentricity or inclination for the planetesimals. This is because our planetesimal population starts from nearly circular and planar orbits. As for the semimajor axis, implantation depends on when Jupiter becomes big enough to be capable of shepherding particles efficiently (Figure \ref{Fignew}). That varies slightly between {\it planetary growth tracks}. But, overall, more than 50\% of the implanted objects come from 6 au $\le a_{planetesimal}^{initial}\le$ 9 au for the cases studied (10 au $\le a_{Jup}^{init} \le$ 25 au).

\begin{figure}[t]
    \centering
    \includegraphics[width=8.cm]{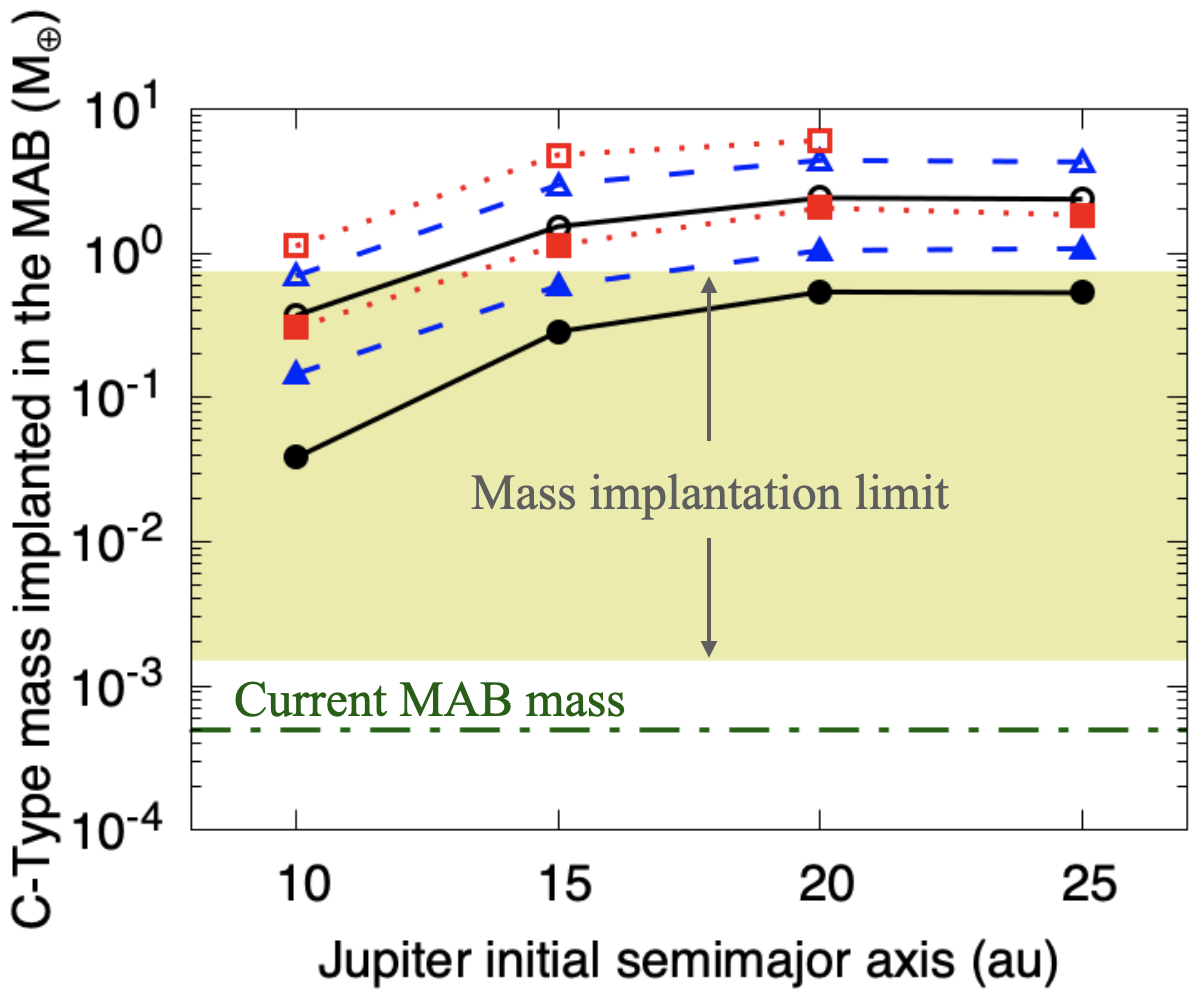}
    \caption{Total mass in C-type planetesimals implanted in the MAB associated with the different {\it planetary growth tracks} and disk masses adopted in Figure \ref{Fig1}, i.e., for Jupiter growing and migrating from 10, 15, 20, and 25 au all the way to 5.2 au. Symbols are the same as in Figure \ref{Fig1}B. The figure also shows, for reference, (i) the current mass of the MAB (green dotted-dashed line; $\approx$ 5$\times$10$^{-4}$ M$_{\oplus}$), and (ii) the maximum implanted amount of mass (shaded yellow; 1.5$\times$10$^{-3}$  M$_{\oplus}$ $\le$ M$_{\rm {Implanted}}$ $\le$ 0.75 M$_{\oplus}$) of C-type planetesimals that could be depleted by models of the giant planet instability followed by depletion due to subsequent chaotic diffusion. Note that the results related to the case 2.0 MMSN, fpl = 0.5 when Jupiter starts at 25 au were suppressed because the very massive disk interior to Jupiter induced planetesimal-driven migration that caused Jupiter’s growth track to diverge from its analytical path (see Section \ref{sec:model}). Therefore, we did not evaluate this case.
    }
    \label{Fig2}
\end{figure}

Figure \ref{Fig2} shows the total amount of C-type mass implanted into the MAB via MMRs by the growing and migrating Jupiter. Figure \ref{Fig2} reports combinations of initial disk mass and Jupiter migration length that would be compatible with current observations (points overlapping with the yellow shaded region). Specifically, Figure \ref{Fig2} shows that Jupiter could have migrated from 15 au if the interior disk mass was $\lesssim$ 3 M$_{\oplus}$ (1.0 MMSN, fpl = 0.1, $a_{Jup}^{init}=$ 15 au, a$_{disk}=$ 5-13 au) or from 10 au if M$_{disk} \lesssim$ 5.64 M$_{\oplus}$ (1.0 MMSN; fpl = 0.5, $a_{Jup}^{init}=$ 10 au, a$_{disk}=$ 5-8 au). 
This is plausible within models where Jupiter would form from a ring of planetesimals around the water-ice line that extends up to 10--15 au \cite[e.g.][see also \cite{Drazkowska2018,Morbidelli2022}]{Izidoro2022}. Alternatively, Jupiter could have migrated from beyond 15 au only if very few planetesimals would have formed between the water-ice line and its original orbit (sub-MMSN cases represented by black dots in Figure \ref{Fig2}; M$_{disk} \lesssim$ 2.44 M$_{\oplus}$ for 0.5 MMSN, fpl = 0.1, $a_{Jup}^{init}=$ 20 au, a$_{disk}=$ 5-18 au; and M$_{disk} \lesssim$ 3.38 M$_{\oplus}$ when 0.5 MMSN, fpl = 0.1, $a_{Jup}^{init}=$ 25 au, a$_{disk}=$ 5-23 au; Figure \ref{Fig1}B). All other scenarios would implant far too much mass into the MAB. That would demand more than 99.9\% depletion. This level of depletion, as previously discussed, only seems possible to achieve by giant planet evolutions that fail in reproducing our current outer solar system architecture \citep{Clement2019}. 

A very large scale migration of Jupiter ($a_{Jup}^{init} \gg$ 15 au) would not be a problem if almost no planetesimals were available inside Jupiter's orbit (i.e., similar to the sub-MMSN cases, black dots, in Figure \ref{Fig2} that we discussed above). Still, given how far from the water-ice line Jupiter would originally be in these cases, that would contradict well-accepted models for planetesimal formation \citep{Drazkowska2017,Drazkowska2018,Lichtenberg2021,Izidoro2022,Morbidelli2022}. In other words, it would imply that planetesimal formation at the water-ice line should be a very inefficient process. Lastly, we should point out that our analysis relies on the fact that absolutely no mass existed in the MAB before implantation \citep{Raymond2017b,Izidoro2022}. We are also not accounting for additional implantation from planetesimals originally beyond Jupiter, e.g., the Jupiter-Saturn zone \citep{Raymond2017a} or farther out \citep{Ribeiro2022}. Therefore, the values reported in Figure \ref{Fig2} should be interpreted as an upper limit on mass implantation. Altogether, this suggests that Jupiter is more likely to have formed within $a\lesssim$ 15 au.

\section{SFD of the implanted and original planetesimal population} \label{sec:sfd}

In our model, we made an assumption that the original SFD of the planetesimal population interior to Jupiter's orbit would follow a cumulative slope $q=$ 5 ($N(>D) \propto D^{-q}$). Although simulations of the streaming instability \citep[e.g.,][]{Youdin2005} and observations from large Kuiper belt objects \citep{Fraser2014} support this cumulative slope, we acknowledge that our assumption is not necessarily correct. For that reason, we now turn our attention to the implications of our choice. Our goal in this section is to understand what the relationship is between the SFD of the MAB implanted population and the SFD of the original planetesimal population interior to Jupiter. 

\begin{figure}[t]
    \centering
    \includegraphics[width=\linewidth]{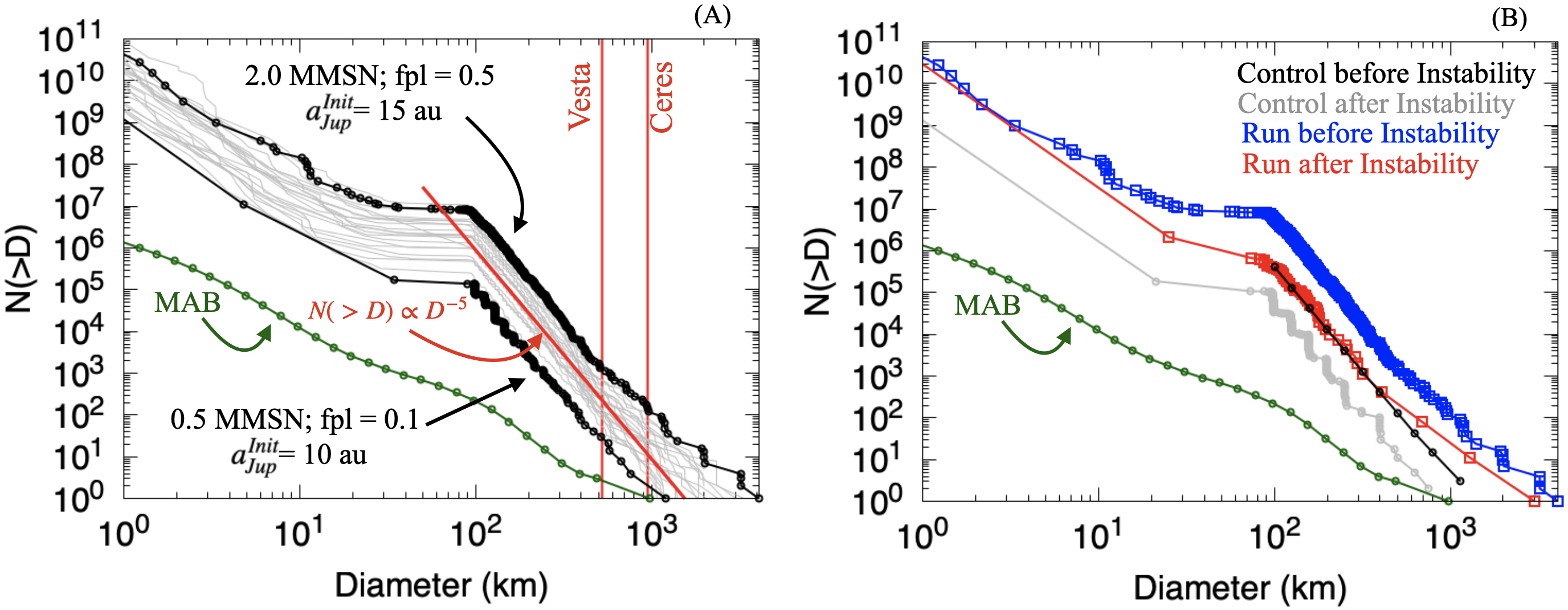}
    \caption{Panel (A) shows the cumulative SFD of the population of C-type planetesimals implanted into the MAB from Figure \ref{Fig2}. Two end-members for the SFD are labeled in black. The intermediate implanted SFDs are shown in gray. They all have the same slope of $N(>D) \propto D^{-5}$ (red oblique line) of the initial planetesimal population for D $>$ 100 km. The current MAB SFD \citep{Bottke2005} is shown in green. Asteroids 1-Ceres and 4-Vesta are also shown for reference (red). Panel (B) reports the results from two experiments where we followed the evolution of the implanted SFDs during depletion by the giant planet instability while still accounting for collisional grinding (see main text for details). The current MAB SFD (green) is repeated for reference.
    }
    \label{Fig3}
\end{figure}

Figure \ref{Fig3}A shows the cumulative SFD of the MAB implanted populations. The slope of the implanted cumulative distributions resembles that of the assumed original population interior to Jupiter, $q=$ 5, at larger sizes (D $>$ 100 km; red oblique line). This happens because large-scale collisional evolution, which may result in large effects on SFD, happens on timescales of tens to hundreds of millions of years \citep{Bottke2005}. Our simulations are performed for 1 Myr total time, and the shepherding process happens on a timescale of a few hundred thousand years. In addition, the level of collisional evolution for low-eccentricity and low-inclination orbits mostly depends on the total mass in planetesimals per disk area. Therefore, collisional evolution of our planetesimal population mostly happens during MMR shepherding because once in MMR, the amount of mass trapped in the narrow resonant area increases. Yet the shepherding process happens too fast to promote enough collisional evolution that would change SFD. That is why only a fragment tail is developed in our simulation, keeping the slope of the distribution of the original population close to its original value for D $>$ 100 km; i.e., larger objects tend to survive.

Small changes in our collisional treatment \citep{Benz1999} would not cause alterations to the SFD predictions. The main changes one can make to the \citet{Benz1999} scaling laws are to (i) increase or decrease the strength of the material that constitutes the planetesimals, making them harder or easier to break, and (ii) increase or decrease the size of the planetesimals where the disruption law would change from the strength to the gravity regime, making it easier to break a particular size of planetesimal (the one at the transition point). These effects were partially studied in \citet{Deienno2020}. In (i) if we increase the material strength, we should observe smaller amounts of generated dust and a less prominent fragment tail. Not much change should be observed for objects with D $>$ 100 km, as the shepherding process is fast. If we were to decrease the material strength, we would eventually make 100 km objects easier to break. Yet, unless we made the planetesimals extremely weak, those with larger diameters would still be strong enough to survive collisions. This would generate a bump (often called a ``knee", an observed feature around D = 100 km in Figure \ref{Fig3}) in the SFD at D $>$ 100 km, e.g., D $\approx$ 200-300 km. We should also observe more dust production and a larger fragment tail. On the other hand, the SFD $N(>D = 200-300~{\rm km})$ in our hypothetical example would retain the original slope. Changes in (ii) would only affect the overall ``wavy" shape (the location of the dips observed in the green curve of Figure \ref{Fig3}) of the SFD after billions of years of collisional evolution, with no effects on the SFD slope \citep{Bottke2015}. 

 A cumulative slope  of $q=$ 5 is much steeper than that of the current MAB \citep[green in Figure \ref{Fig3};][]{Bottke2005}. 
Such a steep slope also results in massive disks leading to the  implantation of planetary embryo-sized objects (D $\approx$ 3000--4000 km) in the MAB, followed by dozens or even hundreds of Ceres-sized bodies and thousands of Vesta-sized objects. Too many of these objects would potentially leave nonobserved dynamical and chemical imprints on the current MAB \citep{OBrien2007,Raymond2009,Zhu2021}. Therefore, if the large majority of C-type asteroids were implanted into the MAB from a population of planetesimals interior to Jupiter, our results indicate that such a population should not have $q=$ 5. This differs from what is suggested by observations of the Kuiper belt and models of the streaming instability. But before we finish with our conclusions, it is important to understand the effects of the giant planet instability and collisional evolution on the implanted SFD.

Let us start evaluating the effects of the giant planet instability happening after protoplanetary gas disk dispersal. Our goal is to understand how MAB orbital excitation, depletion, and  collisional evolution working together would affect the slope of the implanted MAB SFD. Here we did two simple experiments. First, we assumed that the MAB implanted mass was 0.2 M$_{\oplus}$ and $q=$ 5, with $N(>D={\rm 100~km})\approx$ 4.12$\times$10$^5$ and $N(>D={\rm 1000~km})=$ 3 (a case similar to the red oblique line in Figure \ref{Fig3}A, shown in black in Figure \ref{Fig3}B). We also considered the implanted MAB to be dynamically cold, i.e., with circular and planar orbits (assuming that gas drag, which was at play during implantation, was strong enough to damp all implanted orbits) and a semimajor axis between 1.8 au and 3.6 au \citep{Deienno2016,Deienno2018}. In our second experiment, we took the exact orbital distribution for planetesimals within the gray shaded area in the bottom right panel of Figure \ref{Fignew}, which has the implanted SFD from the upper end-member case shown in Figure \ref{Fig3}A (2.0 MMSN, fpl = 0.5, $a_{Jup}^{init}=$ 15 au; blue in Figure \ref{Fig3}B). The giant planet instability is stochastic, and not every simulation reasonably reproduces the outer solar system \citep{Nesvorny2012}. In order to ensure that our giant planets would reproduce the current outer solar system, we used previous successful models of the giant planet instability as a template for the planets' evolution. Using such templates, we can interpolate the orbits of the planets \citep[e.g.,][]{Nesvorny2013}. Specifically, we followed the interpolated giant planet instability from \cite{Deienno2018} \citep[case 3 in][]{Nesvorny2021}. We added the interpolation prescription given by \citet[iSyMBA]{Roig2021} to LIPAD. This implementation is straightforward, since both iSyMBA and LIPAD are written on top of SyMBA \citep{Duncan1998}. The results (Figure \ref{Fig3}B gray and red) showed no changes in the MAB implanted SFD slopes due to depletion via giant planet instability once also accounting for collisional evolution. Only a vertical change (downward) was observed due to the loss of objects. This is because collisional timescales \citep[$t_{coll} \approx \mathcal{O}^{8}$  yrs;][]{Bottke2005} are much larger than the timescales for the giant planet instability \citep[$\Delta t_{inst}<$ 1 Myr;][]{Nesvorny2018a,Deienno2018}. Therefore, with the giant planet instability happening early \citep[$t_{inst}<$ 100 Myr;][]{Nesvorny2018b,Clement2018}, we can conclude that collisional evolution will be important only after MAB depletion. After being depleted by the giant planet instability, the MAB mass will be similar to its current mass \citep{Roig2015,Deienno2016,Deienno2018,Nesvorny2017,Clement2019}. 

With the MAB mass set to its current value,  \cite{Bottke2005} found that the MAB SFD for D $>$ 100 km objects does not change during 4 Gyr of collisional evolution \citep[see also][]{Bottke2015}. \cite{Bottke2005} concluded that the current MAB SFD in this size range is a fossil of its primordial state. We then conclude that, if implanted via Jupiter's inward gas-driven migration, the slope of the implanted SFD indeed reflects that of the initial planetesimal population interior to Jupiter. This suggests that the slope of such a planetesmial population should be similar to the current MAB SFD, and not $q=$ 5, in order to match the current MAB SFD \citep{Bottke2005}. A more important implication is that the SFD of objects formed in the inner portions of the solar system (interior to Jupiter's orbit) may be different than that of the outer solar system.

\section{Implications from generated dust and pebble-sized fragments} \label{sec:nccc}

We now turn our attention to the amount of dust and pebble-sized fragments generated throughout the course of our simulations. Recall that in our setup (Section \ref{sec:model}) planetesimals are assumed to have formed beyond the water-ice line. This is important information, as measurements from meteoritic data suggest that the inner and outer solar system formed from two distinct reservoirs: inner solar system noncarbonaceous chondrites (NCs) and outer solar system carbonaceous chondrites (CCs) \citep[][see also \citet{Leya2008,Budde2016,Dauphas2016,Kruijer2017,Kruijer2020}]{Warren2011}. Subsequent studies and dating from these isotopic groups on iron meteorites demonstrated that the NC and CC reservoirs formed around 0.5--1 Myr after CAI, coexisted, and remained separated for $t_{gap} \approx$ 2--4 Myr \citep[e.g.,][]{Kruijer2017,Bollard2017}. This separation of the NC and CC groups is often called the {\it NC-CC isotopic dichotomy}. Recent results further indicate that CC parent bodies (due to their highly oxidized iron core) formed in a wet environment, i.e., beyond or at the water-ice line, whereas NC parent bodies (with nonoxidized iron cores) formed in a dry environment, i.e., interior to the water-ice line \citep{Bermingham2020,Morbidelli2022}. It has also been measured that the amount of CC material delivered for both Earth and Mars was on the order of a few percent in mass \citep[e.g. 4\% but not exceeding 10\%;][]{Burkhardt2021}.

\begin{figure}[h!]
    \centering
    \includegraphics[width=\linewidth]{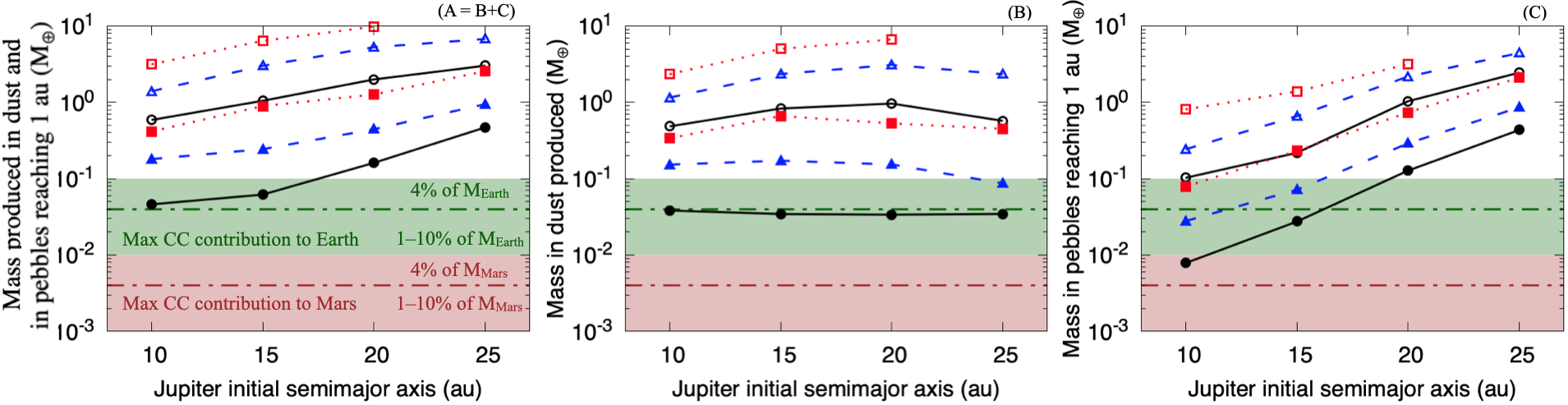}
    \caption{Panel (A):  Total mass of CC dust produced during Jupiter’s migration plus the total amount of mass generated in CC pebble-sized fragments that reached 1 au for the different {\it planetary growth tracks} considered in Figure \ref{Fig1}. Panel (A) is the sum of the contribution shown in panel (B), the total mass of CC dust produced during Jupiter’s migration, and panel (C), the total mass generated in CC pebble-sized fragments that reached 1 au due to fast orbital decay. Green and red shaded areas in panels (A) to (C) demarcate the range of the maximum mass budget contribution that both Earth and Mars (respectively) could receive from outer solar system material based on analysis from meteorite collection, where 4\% (dotted-dashed lines) represents the more probable amount \citep{Burkhardt2021}.
    }
    \label{Fig4}
\end{figure}

Following this rationale, our C-type planetesimals would also present CC isotopic signatures. 
Measuring the amount of dust and pebble-sized fragments generated in CC dust and pebble-sized fragments is then important because these populations can drift and mix with the NC group during the protoplanetary disk lifetime. Such measurements are provided in Figure \ref{Fig4}. Figure \ref{Fig4} shows the values of the mass produced in CC dust and pebbles for all cases from Figure \ref{Fig1}. 
 
Figure \ref{Fig4}A shows that between $\approx$ 0.04 M$_{\oplus}$ and 10 M$_{\oplus}$ in CC dust and pebbles (combined) is produced and transported to the inner solar system during Jupiter's inward migration. Still, pebble accretion efficiency by terrestrial protoplanets is very low \citep[$\epsilon_{acc} \lesssim$ 1\%;][]{Levison2015b,Lambrechts2019}. The numbers reported in Figure \ref{Fig4} then indicate that the amount of mass generated in CC dust and pebbles would likely cause no visible effects on the chemistry of the Earth or Mars. Thus, the {\it NC-CC isotopic dichotomy} would be preserved. On the other hand, it is not clear whether the {\it NC-CC isotopic dichotomy} would remain unchanged in respect to the late-formed chondrites (e.g., ordinary chondrites). That would depend on the amount of NC dust in the inner solar system at the time of late NC chondrite formation \citep[t$_{\rm {chondrite}}\approx$ 2--3 Myr after CAI; e.g,][]{Pape2019}. Dust coagulation models suggest that most, if not all, of the primordial inner solar system NC dust was lost by t $\ll$ 1 Myr, with a fraction of that converted into NC planetesimals  \citep{Drazkowska2017,Lichtenberg2021,Izidoro2022,Morbidelli2022}. Similarly, dynamical models of terrestrial planet formation starting from planetesimal-sized bodies suggest most of those early-formed inner solar system NC planetesimals were consumed by growing protoplanets by t $\lesssim$ 1 Myr after planetesimal formation \citep[t $<$ 2--3 Myr $\approx$ t$_{\rm {chondrite}}$;][]{Levison2015b,Walsh2016,Walsh2019,Deienno2019}. Late-formed NC dust in the inner solar system would then essentially come from collisions among surviving  NC planetesimals, which may not be substantial \citep[M$_{Dust}^{LateNC} \lesssim$ 0.35 M$_{\oplus}$; see Supplementary Figure 8 in][]{Izidoro2022}. This late-formed NC dust would mix with late CC dust produced during Jupiter's inward gas-driven migration (Figure \ref{Fig4}). Dust grains retain the isotopic information from where their parent bodies formed \citep{Spitzer2020}. Therefore, several Earth masses of CC dust (Figure \ref{Fig4}) entering the inner solar system would potentially be enough to make the chondrites found in late-formed planetesimals deviate in their isotope ratios from the observed NC iron meteorites. In this case, the {\it NC-CC isotopic dichotomy} would not be preserved. An alternative solution for this issue is that a preexisting \citep{Brasser2020}, or early-developed \citep{Izidoro2022}, pressure bump was present in the disk around the water-ice line. This pressure bump would potentially prevent CC dust and pebble-sized fragments from penetrating the NC region.

\section{Implications from Jupiter formation time and speed of migration} \label{sec:timing}

Until now, we only focused on the implications of Jupiter inward gas-driven migration to the inner solar system based on the assumption that Jupiter's core formed fast and that the planet migrated from several aus to its current position within 1 Myr after core formation (Section \ref{sec:model}). We now discuss and speculate about some implications that we can infer from our results in case the migration speed was slower or core formation was protracted.

A slower migration of Jupiter would allow for longer interaction of planetesimals transported by MMRs.  In turn, this would necessarily lead to more collisional evolution \citep{Batygin2015,Deienno2020}. The result would be the generation of larger amounts of CC dust and pebble-sized fragments from the time Jupiter starts growing and migrating. 

A lengthier growth timescale of Jupiter via pebble accretion implies that the planet necessarily would have only consumed a fraction of the passing pebbles \citep[e.g.,][]{Lambrechts2014}. That means that the very large flux of CC pebbles that is needed to grow Jupiter's core \citep[which depends on Jupiter's growing distance from the Sun; e.g.,][]{Lambrechts2014,Levison2015a,Bitsch2019} would flood into the inner NC region. This would not only result in the mixing of the two reservoirs, but it would also inevitably deliver large amounts of pebbles with CC signatures to Earth and Mars \citep[e.g., $\approx$ 42\% and 36\%, respectively;][see also \cite{Schiller2018}]{Johansen2021}. Although these large amounts seem plausible once evaluating solely calcium and iron isotopes, they are rather unlikely when accounting for a larger spectrum of isotopic anomalies \citep{Burkhardt2021}. 

Alternatively, in the scenario above, one can assume that the water-ice line would be very efficient in blocking CC dust and pebbles from penetrating the inner solar system \citep[][as briefly discussed at the end of the previous section]{Brasser2020,Izidoro2022}. In this case, the large flux of pebbles passing Jupiter's distant forming core would be transformed into CC planetesimals at the water-ice line. Although this would eventually alleviate the consequences regarding the {\it NC-CC isotopic dichotomy}, it would likely generate a massive planetesimal disk interior to Jupiter. As observed in Figure \ref{Fig2}, this would result in too many C-type planetesimals being later implanted into the MAB. Large embryos would also eventually grow near or at the water-ice line in such a scenario and later be implanted into the MAB. Still, we have no evidence for the existence of planetary embryos in the MAB \citep{OBrien2007,Raymond2009,Clement2019}. 

Finally, but still in respect to the above scenario of Jupiter's lengthier growth, we could assume that pebbles contributing to Jupiter's core formation solely originated from the late infalled (NC) component of the molecular cloud \citep{Nanne2019}. In this case, there would once again be no early CC contribution to the inner solar system. The question then would be whether (i) an efficient water-ice line would transform several Earth masses of not-yet-observed oxidized NC planetesimals interior to Jupiter's forming site \citep{Izidoro2022} that would later be implanted into the MAB (Figure \ref{Fig2}), or (ii) a leaky water-ice line would prompt formation of nonexistent super-Earths in our solar system \citep{Izidoro2022,Morbidelli2022}.  

\section{Conclusions} \label{sec:concl}

In this work, we addressed the influence of Jupiter larger-scale and inward gas-driven migration on the inner solar system. Specifically, we modeled Jupiter's growth and inward gas-driven migration following analytic formulations from {\it planetary growth tracks} \citep[e.g.;][]{Bitsch2015a,Bitsch2015b,Johansen2017,Johansen2019}. We then assumed that a planetesimal disk would exist interior to Jupiter’s original orbit and attributed an initial total mass and size-frequency distribution (SFD) to that disk (see Figure \ref{Fig1} and Section \ref{sec:model}). Using the code LIPAD \citep{Levison2012}, we allowed the planetesimal population to collisionally evolve throughout the entire simulation while also accounting for gas effects (aerodynamic drag, eccentricity, and inclination damping) on planetesimals. We tracked the evolution of the SFD of such planetesimal populations, the amount of mass implanted into the main asteroid belt (MAB), and the amount of dust generated via collisions.

Our results (Figure \ref{Fig2}) show that, unless very little or no mass existed between 5 au and Jupiter's original orbit, it would be difficult to reconcile the current low mass of the MAB with the possibility that Jupiter migrated from distances beyond 15 au. That would imply that the water-ice line (interior to Jupiter's orbit) is very inefficient for prompting planetesimal formation. This is in conflict with well-accepted theories of planetesimal formation \citep[e.g.;][]{Drazkowska2017,Drazkowska2018,Lichtenberg2021,Izidoro2022,Morbidelli2022}. Therefore, our results suggest Jupiter is more likely to have formed at $a \lesssim$ 15 au. This is in close agreement with models of solar system formation from rings of planetesimals \citep{Izidoro2022,Morbidelli2022}.

Our results also suggest that the SFD implanted in the MAB tends to resemble that of the original planetesimal population interior to Jupiter (Figure \ref{Fig3}), and that subsequent mass depletion and  collisional evolution have very minor effects on the implanted SFD slope ($q$; $N(>D) \propto D^{-q}$). Therefore, the SFD of the planetesimal population interior to Jupiter's original orbit that will be implanted into the MAB, should have a cumulative slope similar to the current MAB SFD in order to reproduce the MAB current SFD \citep{Bottke2005,Bottke2015}. A more important implication is that the SFD of objects formed in the inner portions of the solar system (interior to Jupiter's orbit) may be different from that of the outer solar system \citep[which presents $q \approx$ 5;][]{Fraser2014}.

Finally, as our planetesimals are assumed to have formed beyond or at the water-ice line, they were considered to be isotopically akin to carbonaceous chondrites \citep[e.g.,][]{Izidoro2022,Morbidelli2022}. That said, in regard to the {\it NC-CC isotopic dichotomy}  \citep[e.g.;][]{Trinquier2007,Kruijer2017,Kruijer2020,Nanne2019}, our results show that the amount of CC dust produced via planetesimal-planetesimal collisions induced by Jupiter's inward gas-driven migration is likely insufficient to leave measurable chemical signatures on Earth or Mars that deviate from the values found in the literature \citep[see also \citet{Schiller2018,Johansen2021}]{Burkhardt2021}. It is not clear, however, whether such amounts of CC dust produced would be compatible with the isotopic ratio of NC and CC dust found in late-formed NC chondrites \citep[e.g., ordinary chondrites;][]{Kruijer2020,Spitzer2020}. 

\section*{Acknowledgements}
The authors are very thankful for the anonymous reviewer, who provided very constructive comments and suggestions leading to a substantial improvement of this work. The work of R.D. was supported by the NASA Emerging Worlds program, grant 80NSSC21K0387. A.~I. acknowledges support from The Welch Foundation grant No. C-2035-20200401, NASA grant 80NSSC18K0828 (to Rajdeep Dasgupta),   and the Brazilian Federal Agency for Support and Evaluation of Graduate Education (CAPES), in the scope of the Program CAPES-PrInt, process number 88887.310463/2018-00, International Cooperation Project number 3266. A.M. received funding from the European Research Council (ERC) under the European Union’s Horizon 2020 research and innovation program (grant No. 101019380 HolyEarth). D.N. acknowledges support from the NASA Emerging Worlds program. W.F.B.’s work on this paper was supported by the Psyche mission through NASA’s Discovery Program.





\end{document}